# A FAST SEMI-AUTOMATIC METHOD FOR CLASSIFICATION AND COUNTING THE NUMBER AND TYPES OF BLOOD CELLS IN AN IMAGE

*Hamed Sadeghi, Shahram Shirani, David W. Capson*

Department of Electrical and Computer Engineering
McMaster University, Hamilton, Canada
Emails: sadeghh@grads.ece.mcmaster.ca, shirani@mcmaster.ca, capson@mcmaster.ca

## ABSTRACT

A novel and fast semi-automatic method for segmentation, locating and counting blood cells in an image is proposed. In this method, thresholding is used to separate the nucleus from the other parts. We also use Hough transform for circles to locate the center of white cells. Locating and counting of red cells is performed using template matching. We make use of finding local maxima, labeling and mean value computation in order to shrink the areas obtained after applying Hough transform or template matching, to a single pixel as representative of location of each region. The proposed method is very fast and computes the number and location of white cells accurately. It is also capable of locating and counting the red cells with a small error.

## 1. INTRODUCTION

Visual examination of blood images is an important tool for diagnosis, prevention and treatment of patients. Manual operations on these images are very slow and inaccurate. So, there has been a demand for automation of such operations and many computer aided operations have been introduced to detect, classify and measure objects in hematological cytology.

In fact, analysis of blood samples involves the staining of a sample for examination by microscope. Images taken through the microscope can be automatically or semi-automatically processed using computer vision techniques to classify and count the cells. So far several automatic and semi-automatic schemes have been proposed to perform this task [1]~[6]. For example, in [1], a fast and fully automated method has been proposed for segmentation and border identification of different types of cells. [2] proposes a method for segmentation of white blood cell nucleus based on Gram-Schmidt orthogonalization. In [3], we see a method for counting of red cells based on pulse-coupled neural network and auto-wave. [4] proposes a method for differential counting of white cells. It does the separation of nucleus from the other parts using thresholding because of intensity difference between white cell parts and background and cytoplasm separation using a series of morphological erosions and dilations.

In our method for separation of nucleus, we use thresholding like in [4]. We also make use of Hough transform for circles for locating white cells and template matching for locating and counting of red cells. To explain the proposed method with complete details, we consider a sample image and describe the method step by step. As we will see our method is very fast and the only considerable time consuming operation is computing the cross correlation between the original image and the templates. The processing time for our method is approximately 6 seconds per image which is comparable to the process time claimed in [1]. The method proposed in [1] is fully automated but our method is semi-automatic and needs some manual parameter adjustment and optimization. The proposed method is basically a combination of common image processing and computer vision schemes, but the order and the type of the specific schemes used in each stage makes it a novel approach.

This method consists of three parts. In the first part, some preprocessing is performed on the image in order to prepare it for applying the next operators and in the next two parts a number of image processing techniques are used for segmentation and counting of white and red cells.

## 2. PREPROCESSING THE IMAGE

As it can be seen in Fig. 1, the sample original image has low contrast and there are some undesirable periodic lines in it which are especially visible in the right bottom part (orthant IV) of the image. These are common artifacts in these types of images. Specially, the lines will survive after the edge detection process, which we are going to do later, and can make the edge image look very messy. To eliminate the lines, first we should understand what kind of noise they are. They seem to be the effects of a sinusoidal noise on the columns of the image. If we take a look at the absolute value of FFT transform of the image in Fig. 2, we see two bright points in it. It corresponds to the frequency of this single frequency noise. In order to delete this noise, we filter the image with a low-pass Butterworth filter of order 9 and a cutoff frequency of ¼ .

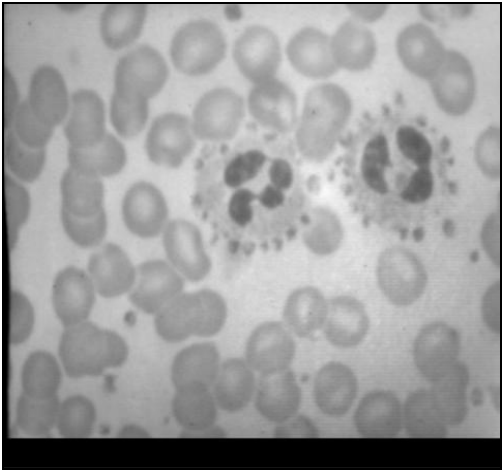

Figure 1: Original sample image

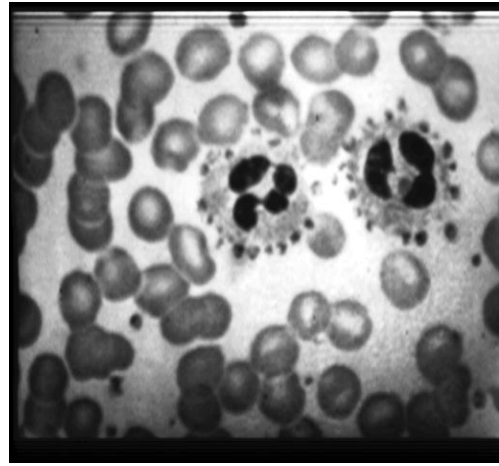

Figure 3: Sample image after lowpass filtering and histogram equalization

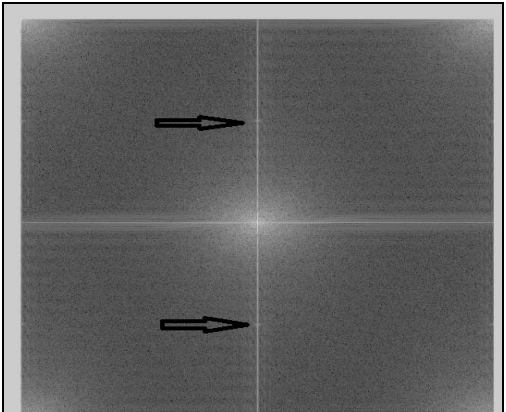

Figure 2: Absolute value of the FFT of the image

To improve the contrast, we do standard histogram equalization on the image. The resulted image can be seen in the Fig. 3. As it is clear in Fig. 3, most of the undesirable lines have been eliminated and the remaining ones are at the top and bottom of the image and will not cause any serious problem in segmentation and counting process of the cells. This image is ready for applying the main method and we will explain the proposed method in the next subsections by performing a series of processes on this image.

### 3. THE PROPOSED METHOD FOR WHITE CELLS

As Fig 3 depicts, nucleus of each white cells has much lower intensity than the other parts of the image. So as stated in [4] by thresholding, we can separate these parts from the other parts. After thresholding and converting the gray-level image to binary using Otsu method [7] for finding the optimal level of thresholding, we come up with an image like what we see in Fig. 4.

In order to count the number of nuclei of white cells (which is equal to the number of white cells) and locate them, first, we label each white region based on 8-connectedness criterion. We exclude some waste parts at the image margins in labeling. For each region, we choose a mid-point. The row number of the mid-point is defined as the mean of maximum and minimum row number of the pixels in the region. Column number is also defined in a similar way. Because each nucleus may have different parts and all of them should be taken into account as one nucleus, we consider the midpoints that are not more than 60 pixels (approximate length of the radius of white cells in this image for its specific scale) far, as one nucleus and assign them the same label in a relabeling process. After relabeling, the number of regions is equal to the number of white cells.

For locating white cells, we find an approximate center pixel for each one. In order to find such a center pixel, we search for the center of circles for circles of the 60 pixels radius in the edge image using Hough transform. To perform such a search, we need a search area for each white cell. To find this area, we make the use of mid-points of the different regions in the white cells. In fact, we find the maximum and minimum row number and set them as the lower and upper rows of the rectangular search region.
We follow the same approach for finding the maximum and minimum column number. After finding the search regions, we extract the edge image using canny edge detector method. We can see the edge image in Fig. 5.
As we said, we try to find the center of nuclei by applying Hough transform for circles on the edge image. We should notice that there may be some fake regions that survived after thresholding that do not belong to any nucleus. These regions can be excluded when we search for centers of white cells because no pixel in them will get votes as representative of a center pixel. So we do not introduce them as a nucleus and subtract their number from the total number of white cells.

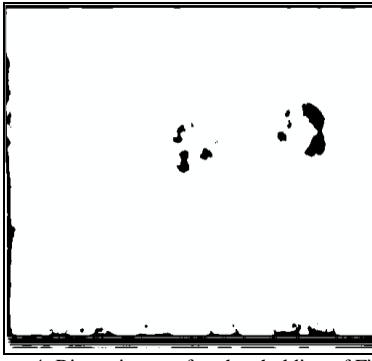
Figure 4: Binary image after thresholding of Fig. 3

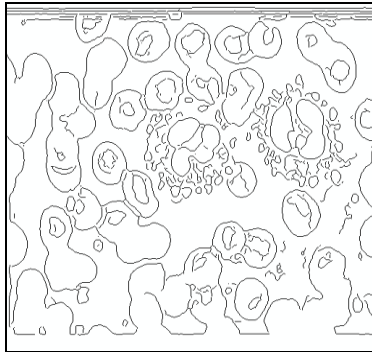
Figure 5: Edge image of the image in Fig. 3.

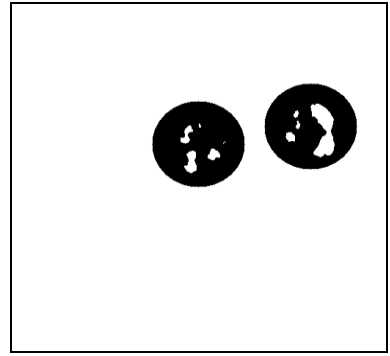
Figure 6: Recognized white cells

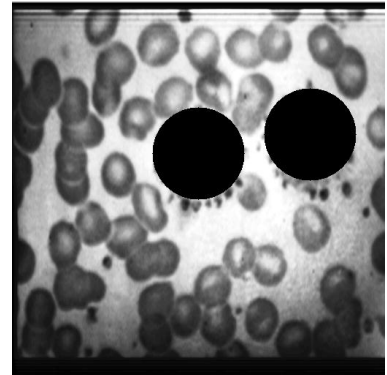
Figure 7: Prepared image for analyzing red cells

In order to find the parts belonging to cytoplasm, we can assign a new color to the parts of located circles that were not recognized as nucleus. We can see the results in Fig. 6. The number of white cells computed to be 2, which is correct.

After recognizing the white cells and determining their location, we can subtract these parts from the preprocessed image (Fig. 3) and obtain a new image that only has red cells. We see such an image in Fig. 7.

## 4. THE PROPOSED METHOD FOR RED CELLS

In order to locate and count the number of red cells in the processed image, we choose five red cells as templates (the numbered cells in Fig. 8) and compute the cross correlation between the processed image and them. Then, we linearly combine the resulted images of the correlation computing process in order to obtain an image in which for each red cell there is a peak value. In order to do this, we should optimize the coefficients. It seems that the red cells which do not considerably intersect with the other parts can be better templates like templates number 36 and 46. So we increase their impact on the resulted image by selecting bigger coefficients for them as we define in (1):

$$Res_0 = Res_1 + Res_{28} + 1.2 \times Res_{36} + Res_{45} + 1.2 \times Res_{46} \quad (1)$$

Here, $Res_i$ is the resulted image after computing the correlation with template number "i" and $Res_0$ is the final image after linear combining. We can see the $Res_0$ image in Fig. 9. In this image, at the location of red cells, we see peak areas rather than peak pixels. In order to count the number of these areas and find their center pixel, we first use thresholding to remove some non-maxima pixels. After that, we find local maximum in each region in order to shrink the peak area more and more. Then we label the shrunken areas and find the number of them which is the number of red cells. For the purpose of locating, we introduce the location of the mid-point of each region as the mean location between all of the pixels in that region.

## 5. RESULTS

If we assign red color to red cells, yellow to cytoplasm, cyan to nucleus and blue to the background, using the proposed method, we will recognize and separate different parts in the image as shown in Fig. 10. Red cells that were used as templates are numbered in the image. The number of recognized red cells is 49, which is close to the actual number (46). We also applied our method to the image depicted in Fig. 11 and the resulted image is depicted in Fig. 12. As we can see in Fig. 12, there are some pixels recognized as nucleus but did not get votes after applying Hough transform. So the obtained number of white cells is a true number which is 2. In this case the recognized number of red cells is 52 which are the same as the actual number.

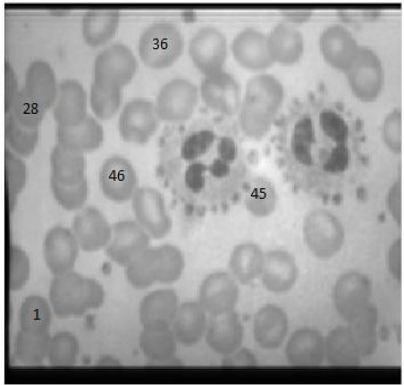

Figure 8: Selected red cells as templates

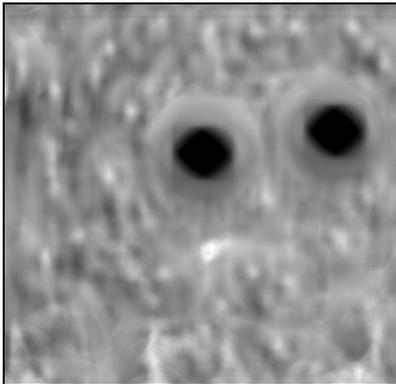

Figure 9: Resulted image of correlation computing process

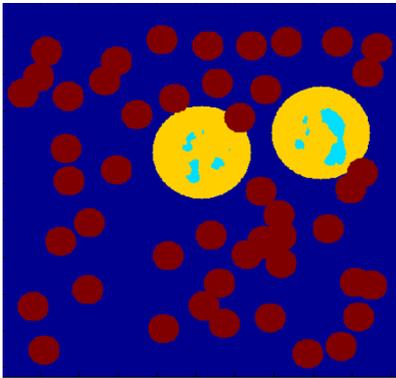

Figure 10: The resulted image after complete segmentation

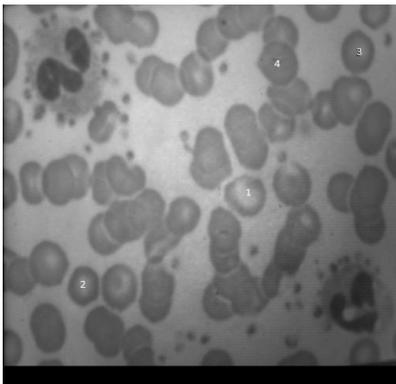

Figure 11: Another sample image and its numbered templates

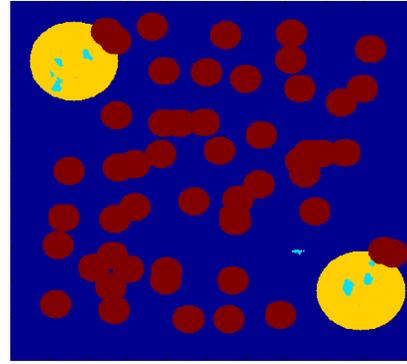

Figure 12: The resulted image after applying the proposed method

## 6. CONCLUSION

We proposed a fast method for segmentation and counting the number of blood cells in an image. As we saw, this method showed a near perfect performance in counting and locating white cells. It was also capable of recognizing artificial low intensity regions caused by poor illumination and removing them from the set of white cells. Furthermore, it could locate and count the number of red cells with a small error.

## 7. REFERENCES

[1] N. Ritter, J. Cooper, "Segmentation and border identification of cells in images of peripheral blood smear slides", thirtieth Australasian computer science conference proceedings, Ballarat Australia, pp. 161-169, 2007.

[2] S.H. Rezatofighi, H. Soltanian-Zadeh, R. Sharifian, R.A. Zoroofi, "A new approach to white blood cell nucleus segmentation based on Gram-Schmidt orthogonalization," icdip, international conference on digital image processing, pp.107-111, 2009.

[3] S. Mao-Jun, W. Zhao-bin, Z. Hong-Juan, M. Yi-de, "A new method for blood cell image segmentation and counting based on PCNN and autowave", ISCCSP, Malta, pp. 6-9, 2008.

[4] F. Sheeba, M. T. T. Hannah, J. J. Mammen, "Segmentation and Reversible Watermarking of Peripheral Blood Smear Images", IEEE fifth international conference on bio-inspired computing: theories and applications (bic-ta), pp. 1373 – 1376, 2010.

[5] N. Theera-Umpon, P. D. Gader, "Training neural networks to count white blood cells via a minimum counting error objective function", 15th international conference on pattern recognition, vol. 2, pp.2299, 2000

[6] S. Nilufar, N. Ray, H. Zhang, "Automatic blood cell classification based on joint histogram based feature and bhattacharya kernel", 42nd asilomar conference on signals, systems and computers, pp. 1915-1918, 2008.

[7] N. Otsu, "A Threshold Selection Method from Gray-Level Histograms," IEEE Transactions on Systems, Man, and Cybernetics, Vol. 9, No. 1, pp. 62-66, 1979.